\documentclass[11pt,conference]{IEEEtran}
\IEEEoverridecommandlockouts
\usepackage{cite}
\usepackage{amsmath,amssymb,amsfonts}
\usepackage{algorithmic}
\usepackage{booktabs}
\usepackage{graphicx}
\usepackage{multirow}
\usepackage{textcomp}
\usepackage{url}
\usepackage{xcolor}
\usepackage[caption=false]{subfig}
\def\BibTeX{{\rm B\kern-.05em{\sc i\kern-.025em b}\kern-.08em
    T\kern-.1667em\lower.7ex\hbox{E}\kern-.125emX}}
\begin{document}

\title{Evaluation of LLMs in Speech is Often Flawed: Test Set Contamination in Large Language Models for Speech Recognition}

\author{
    \IEEEauthorblockN{
        Yuan Tseng*%
        \thanks{*Work done during Yuan's internship at Samsung AI Center Cambridge.} \hspace{2mm}
        Titouan Parcollet%
        \hspace{2mm}
        Rogier van Dalen%
        \hspace{2mm}
        Shucong Zhang%
        \hspace{2mm}
        Sourav Bhattacharya%
    }
    \IEEEauthorblockA{
        \textit{AI Center Cambridge, Samsung, United Kingdom} \\
        \{t.parcollet, r.vandalen, s1.zhang, sourav.b1\}@samsung.com
    }
}

\maketitle

\begin{abstract}
Recent work suggests that large language models (LLMs) can improve performance of speech tasks compared to existing systems.
To support their claims, results on LibriSpeech and Common Voice are often quoted.
However, this work finds that a substantial amount of the LibriSpeech and Common Voice evaluation sets appear in public LLM pretraining corpora.
This calls into question the reliability of findings drawn from these two datasets.
To measure contamination impact, LLMs trained with/without contamination are compared.
A contaminated LLM is more likely to generate test sentences it has seen during training.
Then, speech recognisers based on LLMs are compared.
They show only subtle error rate differences if the LLM is contaminated, but assign significantly higher probabilities to transcriptions seen during LLM training.
Results show that LLM outputs can be biased by tiny amounts of data contamination, highlighting the importance of evaluating LLM-based speech systems with held-out data.
\end{abstract}


\section{Introduction}

Large language models (LLMs) have recently gathered significant amounts of interest due to their strong performance on a wide range of natural language processing (NLP) tasks.
This has led speech researchers to investigate whether LLMs can be similarly useful for speech processing tasks, such as automatic speech recognition (ASR)~\cite{fathullah2024prompting,salsa,10890391,hsu2024let}, speech translation~\cite{cosmic,speechllama}, and ASR error correction~\cite{ma2023can,10389637,gu2024denoising}.

Although some prior work claims that LLM-based speech systems outperform previous systems, their conclusions are often drawn from experiments on standard benchmark corpora.
Such an evaluation methodology overlooks a major concern: LLMs may be inadvertently pretrained on samples in the test set~\cite{contamination1,contamination2,contamination3}, which can lead to inflated scores and overestimation of system capabilities.
For example, Librispeech~\cite{librispeech} and Common Voice~\cite{commonvoice} are two prominent ASR datasets that suffer from this issue.
LibriSpeech consists of audiobook recordings from the Project Gutenberg website\footnote{\url{https://www.gutenberg.org/}}, and many utterances in Common Voice are derived from sentences in Wikipedia pages.
Unfortunately, both Project Gutenberg and Wikipedia are often used for collecting LLM pretraining data, which means that a portion of the LibriSpeech and Common Voice evaluation sets may be seen by LLMs during training.

Throughout this paper, we divide evaluation set sentences into two categories:
\textit{leaked sentences}, which are sentences that are included in the pretraining data, and \textit{non-leaked sentences}, which are not included in the pretraining data.
In this work, we shed light on how many LibriSpeech and Common Voice evaluation transcriptions are leaked, as well as determine how severely this degree of contamination affects model evaluation.

First, we identify various LLM pretraining corpora and models that include LibriSpeech or Common Voice sentences. Through data analysis, we find that the Pile~\cite{pile}, a commonly used text pretraining corpus, is contaminated with nearly \textbf{two-thirds} of each of the four LibriSpeech evaluation sets and \textbf{one-third} of the Common Voice English evaluation sets (Section~\ref{sec:data_contam}).

Second, we train billion-parameter LLMs from scratch, and intentionally contaminate some of them by interspersing LibriSpeech books throughout pretraining. 
Comparing contaminated LLMs against their uncontaminated counterparts shows that data contamination during pretraining makes LLMs significantly more likely to predict leaked sentences (Section~\ref{sec:intentional_contamination}).

Finally, we train LLM-based ASR systems using both uncontaminated and contaminated LLMs (Section~\ref{sec:asr}).
Although speech recogniser error rates do not change much, we find that ASR systems assign significantly higher probabilities to leaked sentences when contaminated LLMs are used, leading to lower perplexities.
This phenomenon highlights how vulnerable LLM-based speech systems are to data contamination, and future work should take care to ensure that the data used for evaluation remains unseen.

\section{Extent of data contamination}%
\label{sec:data_contam}
Although LLM benchmark contamination is a known issue in NLP~\cite{contamination1,contamination2,contamination3}, the topic has not received attention in the speech processing community.
In this section, we highlight that this problem is especially severe for LibriSpeech and Common Voice, as significant portions of their evaluation sets are leaked sentences.
Hence, the various studies that use LLMs for speech tasks evaluated with LibriSpeech or Common Voice~\cite{xu24d_interspeech,yu2024connecting,slamasr,wavllm,jia2024efficient,yang2024ctc} may require revisiting to ensure their conclusions are not affected by test set contamination. 

\subsection{LibriSpeech}
\label{sec:libri}
The LibriSpeech corpus contains approximately 1,000 hours of audiobooks from the Project Gutenberg website.
However, books from Project Gutenberg have also been used in many LLM pretraining corpora, such as the Pile~\cite{pile}, ROOTS~\cite{roots}, RedPajama v1~\cite{redpajama}, Dolma~\cite{dolma}, TxT360~\cite{llm360k2}.
Furthermore, while it has not been revealed what types of data were used to train the later versions of Llama models~\cite{llama2,llama3}, it is known that Project Gutenberg books were used to train Llama 1~\cite{llama1}.

For each utterance in the LibriSpeech dev-clean, dev-other, test-clean, test-other sets, we check for contamination in the Pile through a two-step process.
We first determine the corresponding Project Gutenberg books containing each utterance via metadata provided in LibriSpeech, giving 210 different books in total.
We then search the Project Gutenberg subset of the Pile training set for variants of the full text for each of the 210 books, using an implementation of the MinHash LSH~\cite[Chapter 3]{minhashlsh} algorithm from the \textit{datasketch}\footnote{\url{https://github.com/ekzhu/datasketch}} library.

Through manual verification of candidate pairs, we empirically identified that documents with Jaccard similarity greater than 0.7 are duplicates, while those below this threshold are not.
We note that duplicated books in the Pile may have slight differences from their counterparts in LibriSpeech, possibly caused by text preprocessing errors in different versions of Project Gutenberg, such as \textit{...in coloured silks} versus \textit{...in\ \ silks}.
We still consider these to be duplicates.

In total, these books contain the transcriptions of 6873 out of 11126 utterances in the LibriSpeech dev and test sets, which means that the Pile is effectively contaminated with nearly two-thirds of the LibriSpeech dev and test sets.

\subsection{Common Voice}
Common Voice, as of version 20.0, is a multilingual corpus containing more than 22,000 hours of read speech from 133 languages.
Recordings are gathered through crowdsourcing, where volunteers read sentences from Wikipedia or other user-submitted sources.
However, many LLM pretraining corpora also contain sentences from Wikipedia~\cite{pile,roots,redpajama,dolma}.

For each utterance in the Common Voice English dev and test sets, we use the Wikipedia Search API\footnote{\url{https://www.mediawiki.org/wiki/API:Search}} to look for articles that contain the transcript within the full text.
We then check for contamination in the Pile by matching each article title with documents in the Wikipedia subset of the Pile training set.
We note that the actual degree of contamination may be larger, as we were unable to efficiently search the Pile for Common Voice utterances that have some differences in wording compared to sentences in Wikipedia, such as \textit{For over one hundred...} versus \textit{For more than one hundred...}.

With this process, we find that 10,388 out of 32,796 utterances in the Common Voice dev and test sets appear verbatim in Wikipedia articles that are included in the Pile, meaning that nearly one-third of utterances are seen during LLM pretraining.

\section{Effects of LLM contamination on text generation probabilities}
\label{sec:text_gen}

Although two-thirds of the LibriSpeech evaluation sets are leaked into the Pile, the leaked sentences only comprise a tiny portion of the total training data.
To understand the effects of contamination on text generation probabilities, we compare the probability of generating leaked sentences versus generating non-leaked sentences for open-source LLMs trained on the Pile (Section~\ref{sec:public_llm}), and also intentionally train contaminated LLMs to explore the effects of contamination under different model sizes and different amounts of total pretraining data (Section~\ref{sec:intentional_contamination}).

\vspace{1mm}
\noindent\textbf{Probability calculation:}
We calculate the average word-level negative log-likelihood by normalizing the total cross-entropy loss with $n_{\text{words}}$, the number of words in all sentences $\mathcal{S}$:
\[
    NLL = \frac{- \sum_{s \in \mathcal{S}}\sum_{y_t \in s}{\log p (y_t| y_1, y_2, ... y_{t-1})}}{n_{\text{words}}}, %
\]
where $y_i$ is the $i$th token in the sentence $s$, and $p({y_t})$ is the probability of the LLM predicting $y_t$.
We report the perplexity, the exponentiated average log-likelihood:
\[
    PPL = e^{NLL}
\]

\vspace{1mm}
\noindent\textbf{Statistical significance testing:}
Since the number of tokens constituting leaked LibriSpeech sentences ($3 \times 10^{5}$ tokens) is extremely small compared to the total number of tokens in the Pile ($3 \times 10^{11}$ tokens), the difference between the training data of contaminated and uncontaminated LLMs is minimal.
To assess the statistical significance of our probability calculations, we estimate 95\% confidence intervals via bootstrapping~\cite{bootstrapping,Confidence_Intervals}.
Specifically, we sample with replacement from the original data multiple times to generate 1,000 resampled datasets, from which we calculate the 2.5th and 97.5th percentiles of the average perplexities to estimate the confidence intervals.

\subsection{Does contamination bias open-source LLMs towards generating sentences seen during training?}
\label{sec:public_llm}

The first investigation, which will yield no clear conclusion, is of a series of open-source models, Pythia~\cite{pythia}.
The Pythia models are trained on the Pile~\cite{pile}, a dataset also used for training various other LLMs~\cite{gptneox,opt,pythia,olmo}.
As determined in Section~\ref{sec:libri}, the Pile is contaminated with two-thirds of the LibriSpeech dev and test sets.
To verify if this could cause bias towards generating the LibriSpeech transcriptions that are included in the Pile, we compare the perplexities of leaked and non-leaked LibriSpeech sentences.
Note that Pythia models were trained on document segments without start-of-sentence tokens, hence the reported perplexities do not consider the first token in each sentence.

\begin{figure}[tb]
  \centering
  \includegraphics{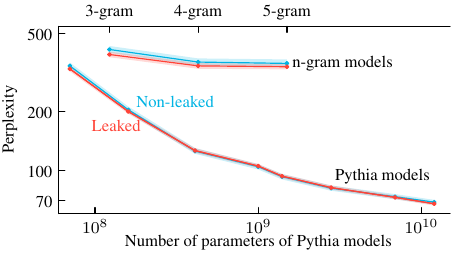}
  \caption{
  Perplexities of leaked and non-leaked sentences calculated by n-gram and open-source Pythia language models.
  We note that leaked and non-leaked sentences have very similar perplexities, which demonstrates the difficulty of measuring the effects of contamination in pretrained models.
  }
  \vspace{-6mm}
  \label{fig:public_llh}
\end{figure}

We restore casing and punctuation to both leaked and non-leaked LibriSpeech transcripts and compare their perplexities. %
The restoration process is performed by matching LibriSpeech transcriptions to the original sentences in Project Gutenberg books and the Pile.
We manually fix all sentences that do not match due to text preprocessing errors mentioned in Section~\ref{sec:libri}.

Figure~\ref{fig:public_llh} compares the perplexities that Pythia models assign to leaked sentences (LibriSpeech dev and test set sentences in the Pile) and non-leaked ones.
Leaked sentences have slightly higher probability of being generated by the Pythia models.
However, we cannot be certain whether this difference in perplexity is truly the result of data contamination, or whether it is that leaked sentences have a different distribution and are more probable.

In order to eliminate this factor, we train uncontaminated n-gram language models on the Librispeech language modeling data\footnote{\url{https://www.openslr.org/11/}}, which is disjoint from the LibriSpeech dev and test sets, hence neither leaked nor non-leaked sentences are included.
To ensure results are comparable with Pythia models, the training data is tokenized with the same tokenizer used for the Pythia models. 
The n-gram results in Figure~\ref{fig:public_llh} suggest that leaked sentences are inherently more likely to be generated.
We conclude that the effects of contamination cannot be easily observed from perplexities calculated by pretrained LLMs, as any differences may reflect inherent differences between the distributions of leaked and non-leaked sentences, rather than bias induced by contamination.
This leads us to conduct additional experiments to better study the effects of contamination.

\subsection{Comparing LLMs trained with/without contamination}
\label{sec:intentional_contamination}

As Section~\ref{sec:public_llm} has shown the difficulty of observing effects of contamination in publicly available models, we take a step further by training Pythia models from scratch, in order to compare the perplexities of models pretrained with and without contamination.

To maintain reasonable computational costs, we train each model on 15B to 60B tokens, as pretraining 1.4B/\nolinebreak[0]2.8B/\nolinebreak[0]6.9B models on 30B tokens (roughly 10\% of the total tokens in the Pile) already requires 4/\nolinebreak[0]7/\nolinebreak[0]15 days of pretraining on 8 A100 GPUs.
We simulate a realistic ratio of leaked data to total pretraining data, by randomly mixing the 82 books containing LibriSpeech sentences not originally in the Pile into the pretraining data of contaminated models.
Hence, leaked sentences refer to LibriSpeech sentences not originally in the Pile from now on.
Both contaminated and uncontaminated models are trained for exactly the same number of steps, to ensure that they are trained on the exact same amount of data.
All other hyperparameters follow the original setups used in~\cite{pythia}.

\begin{figure}[tb]
  \centering
  \subfloat[
  Comparison of LLMs of varying sizes. 
  Contaminated LLMs become more likely to generate leaked sentences, and less likely to generate non-leaked sentences.
  \label{contaminated_llh_sizes}]{
    \includegraphics{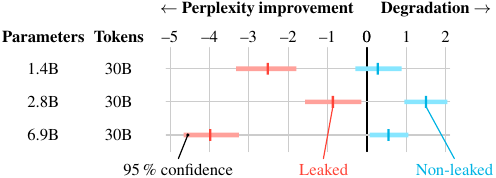}
  }
  \\
  \subfloat[
  Comparison of LLMs trained on varying amounts of data. 
  An increase in training data does not ``dilute'' the effects of contamination; leaked sentences still remain more likely to be generated by contaminated LLMs.
  \label{contaminated_llh_data}]{
    \includegraphics[trim=0 0 -7.5mm 0]{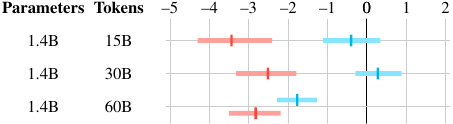}
  }
  \\
  \subfloat[1.4B LLMs that are contaminated once and twice compared against an uncontaminated 1.4B model. All models trained on 30B tokens.
  The more contaminated an LLM is, the higher probability of it generating leaked sentences.
  \label{contaminated_llh_level}]{
    \includegraphics[trim=0 0 -8mm 0]{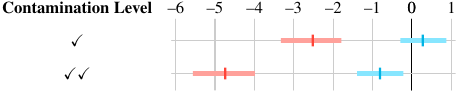}
  }
  \vspace{2mm}
  \caption{
  Comparing perplexity differences between contaminated and uncontaminated LLMs for non-leaked and leaked sentences.
  In the first row of Figure~\ref{contaminated_llh_sizes}, the perplexity of contaminated and uncontaminated LLMs generating leaked sentences is 123.84 and 126.34 respectively, hence a --2.5 difference.
  Generating leaked sentences is always significantly more likely when contaminated LLMs are used.
  }
  \vspace{-6mm}
  \label{fig:contaminated_llh}
\end{figure}

We compare the average perplexity of generating leaked and non-leaked sentences for contaminated and uncontaminated LLMs in Figure~\ref{fig:contaminated_llh}.
We estimate 95\% confidence intervals via bootstrapping.
Results show that contaminated LLMs are consistently more likely to generate leaked sentences, but the probability of generating non-leaked sentences does not increase to a similar extent.
If we examine the effects of contamination across different model sizes and different amounts of pretraining data, we first note that perplexities of leaked sentences improve the most for the largest model (6.9B) we train.
This is consistent with prior work on data memorisation in LLMs: the larger the model size, the more likely it is to memorise its training data~\cite{carlini}.

Furthermore, although increasing the total pretraining data may be expected to dilute effects of contamination, Figure~\ref{contaminated_llh_data} shows that the difference in perplexities of leaked sentences stays roughly the same.
This implies that even if LLMs are trained on an increasing amount of data, as long as the number of leaked sentences remains the same, the effects of contamination may persist.

In addition, we consider the case where test set sentences are included in the LLM pretraining data more than once in Figure~\ref{contaminated_llh_level}.
This can occur when the pretraining data is sampled from multiple sources, with data from specific domains given greater weight than others.
For example, Llama 1 was pretrained on each Project Gutenberg book twice on average~\cite{llama1}.
Therefore, we train an additional 1.4B model on 30B tokens, where each LibriSpeech book occurs in the pretraining data twice.
Similar to the results in Figure~\ref{fig:contaminated_llh}, this resulted in a more pronounced perplexity improvement for leaked sentences (--4.742), and minor improvements for non-leaked sentences (--0.809).

\begin{figure}[tb]
  \centering
  \includegraphics{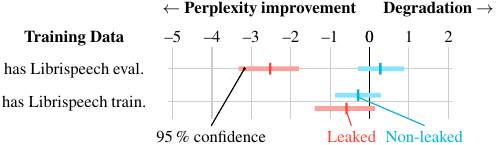}
  \caption{
  Comparison of LLMs contaminated with data from the LibriSpeech eval sets (``leaked'', measuring memorisation), or the LibriSpeech training set (similar distribution but different sentences).
  }
  \vspace{-6mm}
  \label{fig:ablation_llh}
\end{figure}

We also consider an alternative explanation of the improved perplexities of leaked sentences:
Contaminated LLMs are more likely to generate leaked sentences compared to uncontaminated models, because they are exposed to more LibriSpeech books during training.
To verify whether this is the case, we explore another training setup in Figure~\ref{fig:ablation_llh}, where we train a 1.4B model on 30B tokens, with 82 random books containing sentences from the LibriSpeech \textbf{training set}  mixed into the pretraining data.
Compared against the original uncontaminated 1.4B model trained on 30B tokens, we find that this model only has minor improvements on both leaked sentences (--0.583) and non-leaked sentences (--0.288), indicating that the improvements in perplexity cannot be explained by the higher portion of books in the pretraining data of contaminated LLMs.

To summarize, test set contamination during LLM pretraining can inflate the probability of generating leaked sentences while also reducing the probability of generating non-leaked sentences.
This suggests that contamination may not only cause LLMs to exhibit misleadingly high performance on leaked test data, but also degrade performance on actual unseen data.

\section{Effect of using contaminated LLMs for speech recognisers}
\label{sec:asr}

Although results in Section~\ref{sec:intentional_contamination} suggest that even small amounts of contamination can significantly affect the output probabilities of an LLM, it is unknown to what extent this holds true after LLMs are integrated into speech systems.
Therefore, we build LLM-based ASR systems consisting of a speech encoder and an LLM (also inaccurately termed ``decoder-only ASR''), following prior work~\cite{fathullah2024prompting,slamasr}. %
We then explore two experimental settings for speech recognition with LLMs (Section~\ref{sec:asr_settings}). 
We present implementation details in Section~\ref{sec:impl_details}, followed by a discussion of our key findings in Section~\ref{sec:wer}.
\begin{figure}[tb]
  \centering
  \includegraphics{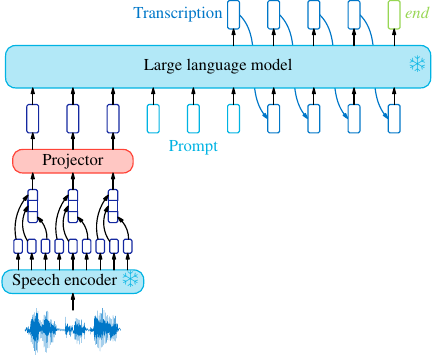}
  \caption{
  A LLM-based speech system for ASR, with speech embeddings stacked then prepended before the text embeddings.
  }
  \vspace{-4mm}
  \label{fig:framework}
\end{figure}

\subsection{Experimental settings}
\label{sec:asr_settings}
We consider two scenarios to understand how speech-conditioned text generation and different downstream data distributions can affect the difference between contaminated and uncontaminated LLM-based speech systems:
\textit{punctuated ASR}, where the LLM is trained on transcriptions in mixed case with punctuation, or \textit{regular ASR}, where transcriptions are in lowercase without punctuation.
Punctuated ASR can be viewed as text generation conditioned on speech embeddings plus text prompts.
On the other hand, the LLMs in regular ASR systems are expected to predict lowercase transcriptions without punctuation, which is very different from the text data distribution on which the LLM is trained.

We train regular ASR systems on 960 hours of data from the LibriSpeech training sets.
For the punctuated ASR systems, we restore punctuation and casing to LibriSpeech train set transcriptions similar to Section~\ref{sec:public_llm}.
Roughly 6400 out of 280,000 training set sentences fail the restoration process, hence we discard the corresponding utterances, and use the remaining 940 hours of data to train the punctuated systems.

\subsection{Implementation details}
\label{sec:impl_details}
We implement our pipeline in the SpeechBrain toolkit~\cite{speechbrainV1}.
As shown in Figure~\ref{fig:framework}, the LLM predicts transcriptions of input speech in an autoregressive fashion, given sequences of speech embeddings obtained from the encoder and an optional text prompt embedding sequence.
Following \cite{slamasr}, we freeze both the speech encoder and the LLM, and only fine-tune parameters of a small projector connecting the two.

We use WavLM Large~\cite{wavlm} as the speech encoder, and stack every 5 embeddings from the encoder before projecting them to the size of the LLM embeddings.
The projector consists of two fully-connected layers, with ReLU activation function.
Each system is trained for up to 100,000 batches using Adam.
The learning rate is warmed up for 1000 steps, until the maximum learning rate of $1\times10^{-4}$ is reached.
As the prompt, we use ``USER: Transcribe speech to text.~\textbackslash{n} ASSISTANT:'', and perform beam search with a beam size of 4.

Unlike \cite{slamasr}, we linearly decay the learning rate to zero and apply SpecAugment during training~\cite{specaugment}, as both improved ASR accuracy for us.
Each experiment is run on a single V100 32GB GPU.
We also use dynamic batching with a maximum batch size of 70 seconds of audio to accelerate training.
Thus, 100,000 batches correspond to about two epochs of training.

\begin{table*}[th]
  \caption{
  Comparing punctuated and regular ASR systems that use uncontaminated or contaminated LLMs pretrained on 30B tokens.
  Each result is averaged over 5 runs with different random seeds.
  Contaminated results that are significantly better are \textbf{bolded}.
  We observe that ASR systems become significantly more likely to generate leaked sentences when using contaminated 6.9B LLMs, but the differences in error rates can be subtle.
  }
  \label{tab:wer}
  \renewcommand{\arraystretch}{1.25}
  \centering
  \setlength\tabcolsep{5.5pt}
  
  \begin{tabular}{ c c cc cc cc cc }
    \toprule
    \multirow{2}{*}[-3pt]{Params.} &
    \multirow{2}{*}[-3pt]{\begin{tabular}{c}Contaminated?\end{tabular}}
    & \multicolumn{2}{c}{dev-clean} & \multicolumn{2}{c}{dev-other} & \multicolumn{2}{c}{test-clean} & \multicolumn{2}{c}{test-other}  \\
    \cmidrule(lr){3-4}
    \cmidrule(lr){5-6}
    \cmidrule(lr){7-8}
    \cmidrule(lr){9-10}
    & & Non-leaked & Leaked        & Non-leaked & Leaked        & Non-leaked & Leaked        & Non-leaked & Leaked        \\
    \midrule
    \multicolumn{9}{l}{Punctuated ASR - Perplexity (conditioned on audio)} \vspace{1mm} \\
    \multirow{2}{*}{1.4B} & {$\times$} &
    1.652 & 1.656 & 1.893 & 1.912 & 1.668 & 1.659 & 1.887 & 1.941 \\
    & {$\checkmark$} &
    1.657 & \textbf{1.651} & 1.888 & 1.918 & 1.668 & 1.654 & 1.885 & 1.939 \\
    \multirow{2}{*}{6.9B} & {$\times$} &
    1.599 & 1.610 & 1.821 & 1.858 & 1.608 & 1.611 & 1.804 & 1.860 \\
    & {$\checkmark$} &
    1.607 & \textbf{1.603} & 1.829 & \textbf{1.834} & 1.619 & \textbf{1.604} & 1.812 & \textbf{1.854} \\
    \midrule
    \multicolumn{9}{l}{Punctuated ASR - CER (includes punctuation errors)} \\
    \addlinespace[1mm]
    \multirow{2}{*}{6.9B} & {$\times$} &
    6.32 & 5.94 & 8.72 & 8.07 & 6.59 & 6.13 & 8.17 & 7.76 \\
    & {$\checkmark$} &
    6.33 & 5.99 & \textbf{8.15} & \textbf{7.67} & 6.55 & 5.80 & 8.24 & 7.67 \\
    \addlinespace[1mm]
    \midrule[\heavyrulewidth]
    \addlinespace[1mm]
    \multicolumn{9}{l}{Regular ASR - Perplexity (conditioned on audio)} \\
    \addlinespace[1mm]
    \multirow{2}{*}{6.9B} & {$\times$} &
    1.146 & 1.172 & 1.308 & 1.356 & 1.153 & 1.161 & 1.309 & 1.347 \\
    & {$\checkmark$} & 
    1.143 & 1.171 & 1.302 & \textbf{1.343} & 1.151 & \textbf{1.156} & 1.304 & \textbf{1.344} \\
    \midrule
    \multicolumn{9}{l}{Regular ASR - WER} \vspace{1mm} \\
    \multirow{2}{*}{6.9B} & {$\times$} &
    3.28    &       3.59    &       7.39    &       6.51    &       3.96    &       3.94    &       6.88    &       6.85 \\
    & {$\checkmark$} & 
    \textbf{2.77}    &       3.77    &       7.01    &       \textbf{5.93}    &       3.61    &       3.50    &       6.89    &       6.97 \\
    \addlinespace[1mm]
    \bottomrule
  \end{tabular}

  \vspace{-4mm}
\end{table*}

\subsection{Experimental results and discussion}
\label{sec:wer}
Table~\ref{tab:wer} shows results on punctuated and regular ASR systems, trained with uncontaminated or contaminated LLMs.
The results are in the form of perplexities (conditional on the audio) and error rates.
Since smaller LLMs produce less noticeable differences (see first two rows and Section~\ref{sec:intentional_contamination}), we focus on 6.9B LLMs.

We see that contaminated 6.9B LLMs still become more likely to generate leaked sentences when used in speech systems, for punctuated and regular ASR systems alike.
In fact, for punctuated ASR, not only does the contaminated 6.9B system always become significantly more likely to generate leaked sentences, it also becomes less likely to generate non-leaked sentences.
Once again, this shows how contamination can not only inflate model performance on leaked data, but also affect how well the model generalises to unseen data.

However, this increased probability of generating leaked sentences is not necessarily reflected in ASR error rates, as the error rates of contaminated systems can worsen even when the perplexities of the same system improve.
Manual inspection of the recognition errors made by contaminated and uncontaminated systems also revealed no obvious differences in their error patterns.
We hypothesise that this is because the increase in perplexity is not large enough to affect the rank ordering of sequences that are taken into consideration by the beam search decoding process.

In summary, results suggest that there is a significant gap in perplexity between contaminated and uncontaminated LLMs, present in both speech-conditioned and unconditional text generation, although the gap proves too subtle for detection through error rates.
Nevertheless, the consistent and significant differences observed in text generation perplexities suggest that data contamination creates subtle but systematic biases.
Proper evaluation of LLMs on held-out data is therefore not only necessary to ensure scientific rigour, but also essential to prevent adverse effects when using LLM-based speech systems for speech translation, dialog systems, or other tasks.

\section{Conclusion}
This work highlights how test set contamination during LLM pretraining can skew output probabilities. %
We point out that LibriSpeech and Common Voice have two-thirds and one-third of their respective evaluation sets included in LLM pretraining datasets.
We also train a series of LLMs from scratch, to show that even tiny amounts of contamination can disproportionately increase the chances of generating leaked sentences.

When contaminated LLMs are used for speech recognition instead of uncontaminated ones, leaked sentences still become more likely to be generated, but to a lesser degree. 
Differences in error rates between contaminated and uncontaminated systems are less consistent but still present.
This suggests using contaminated LLMs for speech tasks can bias evaluation results in subtle ways, emphasizing the importance of evaluating LLM-based speech systems with uncontaminated data.

\vfill\pagebreak
\bibliographystyle{IEEEtran}
\bibliography{mybib}

\begin{thebibliography}{10}
\providecommand{\url}[1]{#1}
\csname url@samestyle\endcsname
\providecommand{\newblock}{\relax}
\providecommand{\bibinfo}[2]{#2}
\providecommand{\BIBentrySTDinterwordspacing}{\spaceskip=0pt\relax}
\providecommand{\BIBentryALTinterwordstretchfactor}{4}
\providecommand{\BIBentryALTinterwordspacing}{\spaceskip=\fontdimen2\font plus
\BIBentryALTinterwordstretchfactor\fontdimen3\font minus \fontdimen4\font\relax}
\providecommand{\BIBforeignlanguage}[2]{{%
\expandafter\ifx\csname l@#1\endcsname\relax
\typeout{** WARNING: IEEEtran.bst: No hyphenation pattern has been}%
\typeout{** loaded for the language `#1'. Using the pattern for}%
\typeout{** the default language instead.}%
\else
\language=\csname l@#1\endcsname
\fi
#2}}
\providecommand{\BIBdecl}{\relax}
\BIBdecl

\bibitem{fathullah2024prompting}
Y.~Fathullah, C.~Wu, E.~Lakomkin, J.~Jia, Y.~Shangguan, K.~Li, J.~Guo, W.~Xiong, J.~Mahadeokar, O.~Kalinli, C.~Fuegen, and M.~Seltzer, ``Prompting large language models with speech recognition abilities,'' in \emph{Proceedings of International Conference on Acoustics, Speech, and Signal Processing}, 2024.

\bibitem{salsa}
A.~Mittal, D.~Prabhu, S.~Sarawagi, and P.~Jyothi, ``Salsa: Speedy asr-llm synchronous aggregation,'' in \emph{Proceedings of Interspeech}, 2024, pp. 3485--3489.

\bibitem{10890391}
T.~Hori, M.~Kocour, A.~Haider, E.~McDermott, and X.~Zhuang, ``Delayed fusion: Integrating large language models into first-pass decoding in end-to-end speech recognition,'' in \emph{Proceedings of International Conference on Acoustics, Speech, and Signal Processing}, 2025.

\bibitem{hsu2024let}
C.-J. Hsu, Y.-C. Chen, F.-T. Liao, P.-C. Ho, Y.-H. Wang, P.-C. Hsu, and D.-s. Shiu, ``Let's fuse step by step: A generative fusion decoding algorithm with llms for multi-modal text recognition,'' \emph{arXiv preprint arXiv:2405.14259}, 2024.

\bibitem{cosmic}
J.~Pan, J.~Wu, Y.~Gaur, S.~Sivasankaran, Z.~Chen, S.~Liu, and J.~Li, ``{COSMIC}: Data efficient instruction-tuning for speech in-context learning,'' \emph{arXiv preprint arXiv:2311.02248}, 2023.

\bibitem{speechllama}
J.~Wu, Y.~Gaur, Z.~Chen, L.~Zhou, Y.~Zhu, T.~Wang, J.~Li, S.~Liu, B.~Ren, L.~Liu, and Y.~Wu, ``On decoder-only architecture for speech-to-text and large language model integration,'' in \emph{Proceedings of IEEE Automatic Speech Recognition and Understanding Workshop}, 2023.

\bibitem{ma2023can}
R.~Ma, M.~Qian, P.~Manakul, M.~Gales, and K.~Knill, ``Can generative large language models perform asr error correction?'' \emph{arXiv preprint arXiv:2307.04172}, 2023.

\bibitem{10389637}
G.~Song, Z.~Wu, G.~Pundak, A.~Chandorkar, K.~Joshi, X.~Velez, D.~Caseiro, B.~Haynor, W.~Wang, N.~Siddhartha, P.~Rondon, and K.~C. Sim, ``Contextual spelling correction with large language models,'' in \emph{Proceedings of IEEE Automatic Speech Recognition and Understanding Workshop}, 2023.

\bibitem{gu2024denoising}
Z.~Gu, T.~Likhomanenko, H.~Bai, E.~McDermott, R.~Collobert, and N.~Jaitly, ``Denoising {LM}: Pushing the limits of error correction models for speech recognition,'' \emph{arXiv preprint arXiv:2405.15216}, 2024.

\bibitem{contamination1}
O.~Sainz, J.~A. Campos, I.~Garc{\'\i}a-Ferrero, J.~Etxaniz, O.~L. de~Lacalle, and E.~Agirre, ``{NLP} evaluation in trouble: On the need to measure {LLM} data contamination for each benchmark,'' in \emph{Findings of Empirical Methods in Natural Language Processing}, 2023.

\bibitem{contamination2}
I.~Magar and R.~Schwartz, ``Data contamination: From memorization to exploitation,'' in \emph{Proceedings of the 60th Annual Meeting of the Association for Computational Linguistics}, 2022.

\bibitem{contamination3}
S.~Balloccu, P.~Schmidtov{\'a}, M.~Lango, and O.~Dusek, ``Leak, cheat, repeat: Data contamination and evaluation malpractices in closed-source {LLM}s,'' in \emph{Proceedings of the 18th Conference of the European Chapter of the Association for Computational Linguistics}, 2024.

\bibitem{librispeech}
V.~Panayotov, G.~Chen, D.~Povey, and S.~Khudanpur, ``Librispeech: An {ASR} corpus based on public domain audio books,'' in \emph{Proceedings of International Conference on Acoustics, Speech, and Signal Processing}, 2015.

\bibitem{commonvoice}
R.~Ardila, M.~Branson, K.~Davis, M.~Kohler, J.~Meyer, M.~Henretty, R.~Morais, L.~Saunders, F.~Tyers, and G.~Weber, ``\BIBforeignlanguage{English}{Common voice: A massively-multilingual speech corpus},'' in \emph{\BIBforeignlanguage{English}{Proceedings of Language Resources and Evaluation Conference}}, 2020.

\bibitem{pile}
L.~Gao, S.~Biderman, S.~Black, L.~Golding, T.~Hoppe, C.~Foster, J.~Phang, H.~He, A.~Thite, N.~Nabeshima, S.~Presser, and C.~Leahy, ``The {Pile}: An {800GB} dataset of diverse text for language modeling,'' \emph{arXiv preprint arXiv:2101.00027}, 2020.

\bibitem{xu24d_interspeech}
Y.~Xu, S.-X. Zhang, J.~Yu, Z.~Wu, and D.~Yu, ``Comparing discrete and continuous space llms for speech recognition,'' in \emph{Proceedings of Interspeech}, 2024.

\bibitem{yu2024connecting}
W.~Yu, C.~Tang, G.~Sun, X.~Chen, T.~Tan, W.~Li, L.~Lu, Z.~Ma, and C.~Zhang, ``Connecting speech encoder and large language model for {ASR},'' in \emph{Proceedings of International Conference on Acoustics, Speech, and Signal Processing}, 2024.

\bibitem{slamasr}
Z.~Ma, G.~Yang, Y.~Yang, Z.~Gao, J.~Wang, Z.~Du, F.~Yu, Q.~Chen, S.~Zheng, S.~Zhang, and X.~Chen, ``An embarrassingly simple approach for {LLM} with strong {ASR} capacity,'' \emph{arXiv preprint arXiv:2402.08846}, 2024.

\bibitem{wavllm}
S.~Hu, L.~Zhou, S.~Liu, S.~Chen, L.~Meng, H.~Hao, J.~Pan, X.~Liu, J.~Li, S.~Sivasankaran, L.~Liu, and F.~Wei, ``{WavLLM}: Towards robust and adaptive speech large language model,'' \emph{arXiv preprint arXiv:2404.00656}, 2024.

\bibitem{jia2024efficient}
J.~Jia, G.~Keren, W.~Zhou, E.~Lakomkin, X.~Zhang, C.~Wu, F.~Seide, J.~Mahadeokar, and O.~Kalinli, ``Efficient streaming llm for speech recognition,'' \emph{arXiv preprint arXiv:2410.03752}, 2024.

\bibitem{yang2024ctc}
G.~Yang, Z.~Ma, Z.~Gao, S.~Zhang, and X.~Chen, ``Ctc-assisted llm-based contextual asr,'' in \emph{Proceedings of Spoken Language Technology Workshop}, 2024, pp. 126--131.

\bibitem{roots}
H.~Lauren{\c{c}}on, L.~Saulnier, T.~Wang, C.~Akiki, A.~V. del Moral, T.~L. Scao, L.~V. Werra, C.~Mou, E.~G. Ponferrada, H.~Nguyen, J.~Frohberg, M.~{\v{S}}a{\v{s}}ko, Q.~Lhoest, A.~McMillan-Major, G.~Dupont, S.~Biderman, A.~Rogers, L.~B. allal, F.~D. Toni, G.~Pistilli, O.~Nguyen, S.~Nikpoor, M.~Masoud, P.~Colombo, J.~de~la Rosa, P.~Villegas, T.~Thrush, S.~Longpre, S.~Nagel, L.~Weber, M.~R. Mu{\~n}oz, J.~Zhu, D.~V. Strien, Z.~Alyafeai, K.~Almubarak, V.~M. Chien, I.~Gonzalez-Dios, A.~Soroa, K.~Lo, M.~Dey, P.~O. Suarez, A.~Gokaslan, S.~Bose, D.~I. Adelani, L.~Phan, H.~Tran, I.~Yu, S.~Pai, J.~Chim, V.~Lepercq, S.~Ilic, M.~Mitchell, S.~Luccioni, and Y.~Jernite, ``The bigscience {ROOTS} corpus: A 1.6{TB} composite multilingual dataset,'' in \emph{Proceedings of the Neural Information Processing Systems Track on Datasets and Benchmarks}, 2022.

\bibitem{redpajama}
\BIBentryALTinterwordspacing
T.~Computer, ``{RedPajama}: an open dataset for training large language models,'' 2023. [Online]. Available: \url{https://github.com/togethercomputer/RedPajama-Data}
\BIBentrySTDinterwordspacing

\bibitem{dolma}
L.~Soldaini, R.~Kinney, A.~Bhagia, D.~Schwenk, D.~Atkinson, R.~Authur, B.~Bogin, K.~Chandu, J.~Dumas, Y.~Elazar, V.~Hofmann, A.~Jha, S.~Kumar, L.~Lucy, X.~Lyu, N.~Lambert, I.~Magnusson, J.~Morrison, N.~Muennighoff, A.~Naik, C.~Nam, M.~Peters, A.~Ravichander, K.~Richardson, Z.~Shen, E.~Strubell, N.~Subramani, O.~Tafjord, E.~Walsh, L.~Zettlemoyer, N.~Smith, H.~Hajishirzi, I.~Beltagy, D.~Groeneveld, J.~Dodge, and K.~Lo, ``Dolma: an open corpus of three trillion tokens for language model pretraining research,'' in \emph{Proceedings of the 62nd Annual Meeting of the Association for Computational Linguistics}, 2024.

\bibitem{llm360k2}
Z.~Liu, B.~Tan, H.~Wang, W.~Neiswanger, T.~Tao, H.~Li, F.~Koto, Y.~Wang, S.~Sun, O.~Pangarkar, R.~Fan, Y.~Gu, V.~Miller, L.~Ma, L.~Tang, N.~Ranjan, Y.~Zhuang, G.~He, R.~Wang, M.~Deng, R.~Algayres, Y.~Li, Z.~Shen, P.~Nakov, and E.~Xing, ``{LLM360} {K2}: Building a {65B} 360-open-source large language model from scratch,'' \emph{arXiv preprint arXiv:2501.07124}, 2025.

\bibitem{llama2}
H.~Touvron, L.~Martin, K.~Stone, P.~Albert, A.~Almahairi, Y.~Babaei, N.~Bashlykov, S.~Batra, P.~Bhargava, S.~Bhosale, D.~Bikel, L.~Blecher, C.~C. Ferrer, M.~Chen, G.~Cucurull, D.~Esiobu, J.~Fernandes, J.~Fu, W.~Fu, B.~Fuller, C.~Gao, V.~Goswami, N.~Goyal, A.~Hartshorn, S.~Hosseini, R.~Hou, H.~Inan, M.~Kardas, V.~Kerkez, M.~Khabsa, I.~Kloumann, A.~Korenev, P.~S. Koura, M.-A. Lachaux, T.~Lavril, J.~Lee, D.~Liskovich, Y.~Lu, Y.~Mao, X.~Martinet, T.~Mihaylov, P.~Mishra, I.~Molybog, Y.~Nie, A.~Poulton, J.~Reizenstein, R.~Rungta, K.~Saladi, A.~Schelten, R.~Silva, E.~M. Smith, R.~Subramanian, X.~E. Tan, B.~Tang, R.~Taylor, A.~Williams, J.~X. Kuan, P.~Xu, Z.~Yan, I.~Zarov, Y.~Zhang, A.~Fan, M.~Kambadur, S.~Narang, A.~Rodriguez, R.~Stojnic, S.~Edunov, and T.~Scialom, ``Llama 2: Open foundation and fine-tuned chat models,'' \emph{arXiv preprint arXiv:2307.09288}, 2023.

\bibitem{llama3}
A.~Grattafiori, A.~Dubey, A.~Jauhri, A.~Pandey, A.~Kadian, A.~Al-Dahle, A.~Letman, A.~Mathur, A.~Schelten, A.~Vaughan, A.~Yang, A.~Fan, A.~Goyal, A.~Hartshorn, A.~Yang, A.~Mitra, A.~Sravankumar, A.~Korenev, A.~Hinsvark, A.~Rao, A.~Zhang, A.~Rodriguez, A.~Gregerson, A.~Spataru, B.~Roziere, B.~Biron, B.~Tang, B.~Chern, C.~Caucheteux, C.~Nayak, C.~Bi, C.~Marra, C.~McConnell, C.~Keller, C.~Touret, C.~Wu, C.~Wong, C.~C. Ferrer, C.~Nikolaidis, D.~Allonsius, D.~Song, D.~Pintz, D.~Livshits, D.~Wyatt, D.~Esiobu, D.~Choudhary, D.~Mahajan, D.~Garcia-Olano, D.~Perino, D.~Hupkes, E.~Lakomkin, E.~AlBadawy, E.~Lobanova, E.~Dinan, E.~M. Smith, F.~Radenovic, F.~Guzmán, F.~Zhang, G.~Synnaeve, G.~Lee, G.~L. Anderson, G.~Thattai, G.~Nail, G.~Mialon, G.~Pang, G.~Cucurell, H.~Nguyen, H.~Korevaar, H.~Xu, H.~Touvron, I.~Zarov, I.~A. Ibarra, I.~Kloumann, I.~Misra, I.~Evtimov, J.~Zhang, J.~Copet, J.~Lee, J.~Geffert, J.~Vranes, J.~Park, J.~Mahadeokar, J.~Shah, J.~van~der Linde, J.~Billock, J.~Hong, J.~Lee, J.~Fu, J.~Chi, J.~Huang,
  J.~Liu, J.~Wang, J.~Yu, J.~Bitton, J.~Spisak, J.~Park, J.~Rocca, J.~Johnstun, J.~Saxe, J.~Jia \emph{et~al.}, ``The {Llama} 3 herd of models,'' \emph{arXiv preprint arXiv:2407.21783}, 2024.

\bibitem{llama1}
H.~Touvron, T.~Lavril, G.~Izacard, X.~Martinet, M.-A. Lachaux, T.~Lacroix, B.~Rozière, N.~Goyal, E.~Hambro, F.~Azhar, A.~Rodriguez, A.~Joulin, E.~Grave, and G.~Lample, ``{LL}a{MA}: Open and efficient foundation language models,'' \emph{arXiv preprint arXiv:2302.13971}, 2023.

\bibitem{minhashlsh}
J.~Leskovec, A.~Rajaraman, and J.~D. Ullman, \emph{Mining of Massive Datasets}, 2nd~ed.\hskip 1em plus 0.5em minus 0.4em\relax Cambridge University Press, 2014.

\bibitem{bootstrapping}
M.~Keller, S.~Bengio, and S.~Wong, ``Benchmarking non-parametric statistical tests,'' 2005.

\bibitem{Confidence_Intervals}
\BIBentryALTinterwordspacing
L.~Ferrer and P.~Riera, ``Confidence intervals for evaluation in machine learning.'' [Online]. Available: \url{https://github.com/luferrer/ConfidenceIntervals}
\BIBentrySTDinterwordspacing

\bibitem{pythia}
S.~Biderman, H.~Schoelkopf, Q.~G. Anthony, H.~Bradley, K.~O'Brien, E.~Hallahan, M.~A. Khan, S.~Purohit, U.~S. Prashanth, E.~Raff, A.~Skowron, L.~Sutawika, and O.~Van Der~Wal, ``Pythia: A suite for analyzing large language models across training and scaling,'' in \emph{Proceedings of International Conference on Machine Learning}, vol. 202, 2023, pp. 2397--2430.

\bibitem{gptneox}
S.~Black, S.~Biderman, E.~Hallahan, Q.~Anthony, L.~Gao, L.~Golding, H.~He, C.~Leahy, K.~McDonell, J.~Phang, M.~Pieler, U.~S. Prashanth, S.~Purohit, L.~Reynolds, J.~Tow, B.~Wang, and S.~Weinbach, ``{GPT}-{N}eo{X}-20{B}: An open-source autoregressive language model,'' in \emph{Proceedings of BigScience Episode {\#}5 -- Workshop on Challenges {\&} Perspectives in Creating Large Language Models}, 2022, pp. 95--136.

\bibitem{opt}
S.~Zhang, S.~Roller, N.~Goyal, M.~Artetxe, M.~Chen, S.~Chen, C.~Dewan, M.~Diab, X.~Li, X.~V. Lin, T.~Mihaylov, M.~Ott, S.~Shleifer, K.~Shuster, D.~Simig, P.~S. Koura, A.~Sridhar, T.~Wang, and L.~Zettlemoyer, ``{OPT}: Open pre-trained transformer language models,'' \emph{arXiv preprint arXiv:2205.01068}, 2022.

\bibitem{olmo}
D.~Groeneveld, I.~Beltagy, E.~Walsh, A.~Bhagia, R.~Kinney, O.~Tafjord, A.~Jha, H.~Ivison, I.~Magnusson, Y.~Wang, S.~Arora, D.~Atkinson, R.~Authur, K.~Chandu, A.~Cohan, J.~Dumas, Y.~Elazar, Y.~Gu, J.~Hessel, T.~Khot, W.~Merrill, J.~Morrison, N.~Muennighoff, A.~Naik, C.~Nam, M.~Peters, V.~Pyatkin, A.~Ravichander, D.~Schwenk, S.~Shah, W.~Smith, E.~Strubell, N.~Subramani, M.~Wortsman, P.~Dasigi, N.~Lambert, K.~Richardson, L.~Zettlemoyer, J.~Dodge, K.~Lo, L.~Soldaini, N.~Smith, and H.~Hajishirzi, ``{OLM}o: Accelerating the science of language models,'' in \emph{Proceedings of the 62nd Annual Meeting of the Association for Computational Linguistics}, L.-W. Ku, A.~Martins, and V.~Srikumar, Eds., 2024, pp. 15\,789--15\,809.

\bibitem{carlini}
N.~Carlini, C.~Liu, {\'U}.~Erlingsson, J.~Kos, and D.~Song, ``The secret sharer: Evaluating and testing unintended memorization in neural networks,'' in \emph{Proceedings of the 28th USENIX Conference on Security Symposium}, 2019, pp. 267--284.

\bibitem{speechbrainV1}
M.~Ravanelli, T.~Parcollet, A.~Moumen, S.~de~Langen, C.~Subakan, P.~Plantinga, Y.~Wang, P.~Mousavi, L.~D. Libera, A.~Ploujnikov, F.~Paissan, D.~Borra, S.~Zaiem, Z.~Zhao, S.~Zhang, G.~Karakasidis, S.-L. Yeh, P.~Champion, A.~Rouhe, R.~Braun, F.~Mai, J.~Zuluaga-Gomez, S.~M. Mousavi, A.~Nautsch, H.~Nguyen, X.~Liu, S.~Sagar, J.~Duret, S.~Mdhaffar, G.~Laperri{{\`e}}re, M.~Rouvier, R.~D. Mori, and Y.~Est{{\`e}}ve, ``Open-source conversational {AI} with {SpeechBrain} 1.0,'' \emph{Journal of Machine Learning Research}, vol.~25, no. 333, pp. 1--11, 2024.

\bibitem{wavlm}
S.~Chen, C.~Wang, Z.~Chen, Y.~Wu, S.~Liu, Z.~Chen, J.~Li, N.~Kanda, T.~Yoshioka, X.~Xiao, J.~Wu, L.~Zhou, S.~Ren, Y.~Qian, Y.~Qian, J.~Wu, M.~Zeng, X.~Yu, and F.~Wei, ``Wav{LM}: Large-scale self-supervised pre-training for full stack speech processing,'' \emph{IEEE Journal of Selected Topics in Signal Processing}, vol.~16, no.~6, pp. 1505--1518, 2022.

\bibitem{specaugment}
D.~S. Park, W.~Chan, Y.~Zhang, C.-C. Chiu, B.~Zoph, E.~D. Cubuk, and Q.~V. Le, ``Spec{A}ugment: A simple data augmentation method for automatic speech recognition,'' in \emph{Proceedings of Interspeech}, 2019, pp. 2613--2617.

\end{thebibliography}

\end{document}